\documentclass[aps,prl]{revtex4}%
\usepackage{amsmath}
\usepackage{graphicx}
\usepackage{amsfonts}
\usepackage{amssymb}%
\begin{document}
\preprint{ }
\title{Comment on ``Fr\"{o}hlich Mass in GaAs-Based Structures''}
\author{S. N. Klimin and J. T. Devreese}
\affiliation{Theoretische Fysica van de Vaste Stoffen (TFVS), Universiteit Antwerpen,
B-2610 Antwerpen, Belgium}
\date{January 24, 2006}
\startpage{1}
\endpage{5}
\maketitle

In a recent Letter \cite{Faugeras2004}, Faugeras et al. investigate the
cyclotron resonance (CR) spectra of doped GaAs quantum well structures. From
an analysis of their data these authors question the validity of the concept
of the Fr\"{o}hlich interaction and the polaron mass in a real system. It is
stated in Ref. \cite{Faugeras2004}, that Fr\"{o}hlich's polaron theory would
predict a resonant magnetopolaron coupling near the longitudinal optical
phonon frequency $\omega_{\mathrm{LO}}$. Because of the absence of
anticrossing at $\omega\approx\omega_{\mathrm{LO}}$ in the experimental data
\cite{Faugeras2004} for a GaAs/AlAs quantum well, they conclude that the CR
spectra ``do not show any sign of interaction related to the Fr\"{o}hlich
coupling'', and therefore, that ``the concept of the polaronic mass, related
to this interaction, is no longer effective'' (Ref. \cite{Faugeras2004}, p.
4). The experimental evidence for the interaction of the electrons with some
modes with frequencies close to the transverse optical phonon frequency
$\omega_{\mathrm{TO}}$ is attributed in Ref. \cite{Faugeras2004} to the
deformation potential. However, the deformation potential, induced by the
optical phonons near the center of the Brillouin zone in cubic semiconductors,
has only off-diagonal matrix elements involving the heavy and light hole
states \cite{BP}. Since the optical conductivity in the experiment
\cite{Faugeras2004} is due to band electrons, the anticrossing near
$\omega_{\mathrm{TO}}$ cannot be interpreted in terms of the deformation
electron-phonon interaction. Furthermore, the phenomenological dielectric
model of Ref. \cite{Faugeras2004} does not predict the anticrossing near
$\omega_{\mathrm{TO}}$.

In contrast to the statement that ``there is no quantum mechanical treatment
of the hybrid modes'' (\cite{Faugeras2004}, p. 1), such a treatment has been
performed in Ref. \cite{CRQW}. In that paper we demonstrated \cite{CRQW} that
in a quantum well with sufficiently high electron density, mixing of bulk-like
and interface polar optical phonons with intrasubband plasmons leads to the
appearance of hybrid magnetoplasmon-phonon modes, which interact with the
electrons. This interaction is the renormalized Fr\"{o}hlich interaction. One
of the frequencies of the hybrid modes, which is close to $\omega
_{\mathrm{TO}}$ in GaAs, provides a dominant contribution to the CR spectrum.
The resulting peak positions, the frequencies of the hybrid
magnetoplasmon-phonon modes as well as the magnitude of the splitting of the
CR peaks calculated in Ref. \cite{CRQW} are in good agreement with the
experimental data \cite{Poulter2001}.

The resonant magnetopolaron coupling also appears in the recent measurements
of the CR spectra for a 13 nm wide GaAs quantum well \cite{Faugeras2004}.
Owing to this magnetopolaron effect, the CR peaks split near $\omega
_{\mathrm{TO}}$ and also change their positions with respect to those obtained
without electron-phonon interaction. These splitting and shift of the CR peaks
near the TO-phonon frequency predicted in Ref. \cite{CRQW} on the basis of the
Fr\"{o}hlich interaction are clearly seen in Fig. 1. The theoretical peak
positions of the CR spectra calculated within the many-polaron approach of
Ref. \cite{CRQW} compare well with the experimental data \cite{Faugeras2004},
as distinct from the CR energies calculated without electron-phonon
interaction, which show no particular features in the region of the
optical-phonon frequencies.

In conclusion, in contrast to the statement of Ref. \cite{Faugeras2004} that
the concept of the Fr\"{o}hlich polaron mass has to be reexamined in a real
material, we conclude that this concept is valid and even necessary to
interpret the CR spectra of quantum wells, e. g., those studied in Ref.
\cite{Faugeras2004}.

The authors acknowledge discussions with V. M. Fomin and F. Brosens. This work
has been supported by the GOA BOF UA 2000, IUAP, FWO-V projects G.0274.01N,
G.0435.03, the WOG WO.025.99 (Belgium) and the European Commission GROWTH
Programme, NANOMAT project, contract No. G5RD-CT-2001-00545.

\newpage%

\begin{figure}
[ptbh]
\begin{center}
\includegraphics[
height=3.3235in,
width=4.4218in
]%
{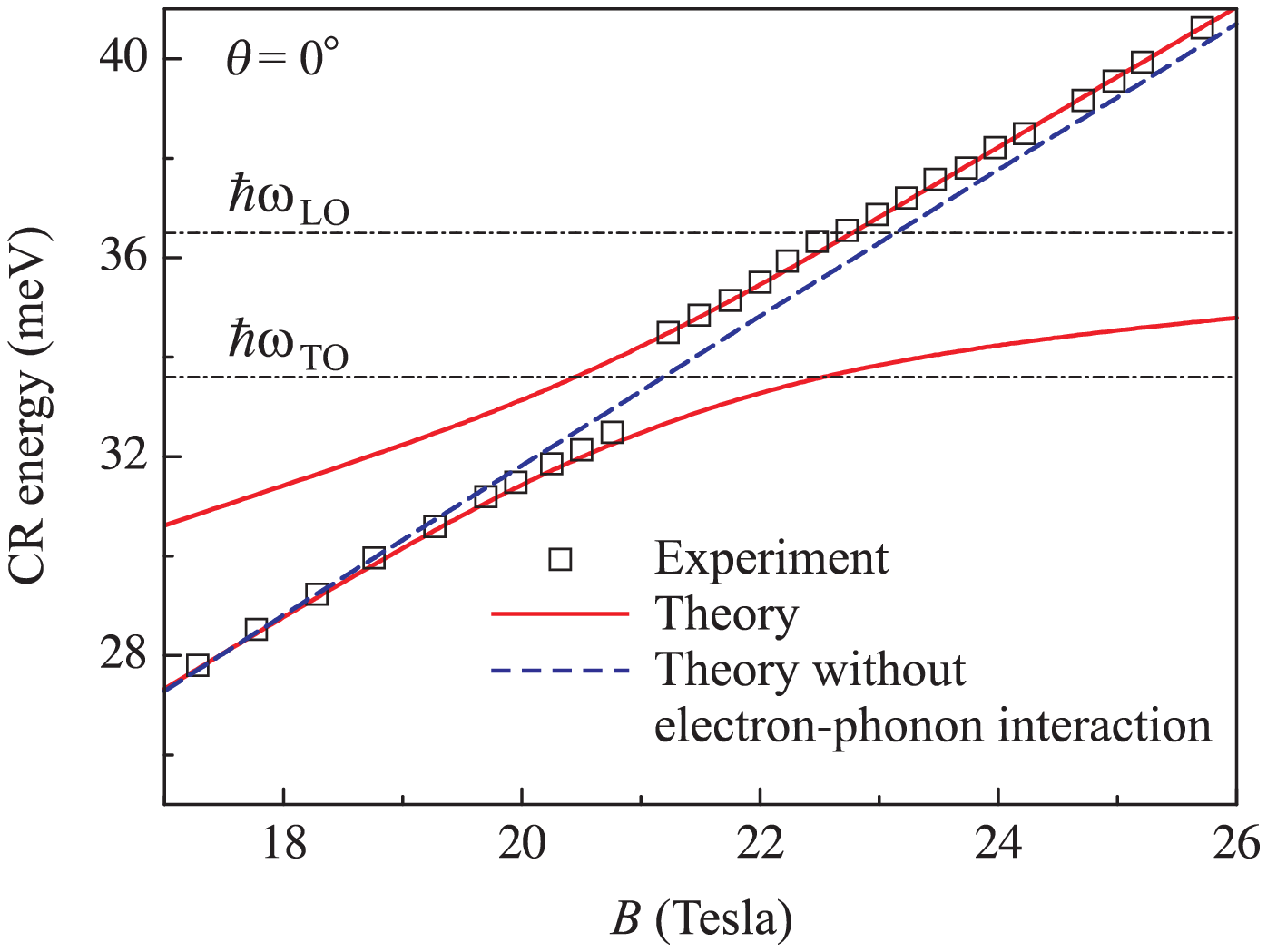}%
\end{center}
\end{figure}

\bigskip

Fig. 1. (Color online) Experimental \cite{Faugeras2004} (squares) and
theoretical (solid curves) CR energies of a 13 nm width GaAs/AlAs quantum well
with the electron density $n_{0}=7\times10^{11}$ cm$^{-2}$ in a perpendicular
magnetic field. The dashed curve represents CR energies calculated without
electron-phonon interaction.

\end{document}